\documentstyle[preprint,aps]{revtex}
\def\lla{\left\langle}
\def\rra{\right\rangle}

\begin{document}
\title{Radiatively Induced Neutrino Majorana Masses and Oscillation}
\author{Darwin Chang$^{a,b}$ \& A. Zee$^{a,c}$}
\address{$^{a}$National Center for Theoretical Sciences, Hsinchu, Taiwan,
30043, ROC\\
$^{b}$Physics Department, National Tsing-Hua University, Hsinchu,
Taiwan, 30043, ROC\\
$^{c}$Institute for Theoretical Physics, University of California,\\
Santa Barbara, California 93106, USA\\
chang@phys.nthu.edu.tw, zee@itp.ucsb.edu}
\date{\today }
\maketitle

\begin{abstract}
We review and remark on models of radiatively induced neutrino Majorana
masses and oscillations. It is pointed out that while the models are capable
of accounting for the observed solar and atmospheric neutrino
oscillation,
some of them can also induce neutrinoless double beta decay and $\mu^-$ --
$e^+$ conversion in nuclei large enough to
be potentially observable in the near future.
\end{abstract}

\bigskip

With the recent results of the SuperKamiokande experiments and
other experiments\cite{superk}, the field of neutrino oscillation and masses
has moved from idle speculation to hard science. Let us ask what is the
simplest modification one can make to the standard model in order to have
neutrino oscillation and masses. Almost twenty years ago, one of us asked
precisely this question and arrived at a model of lepton number violation
and neutrino Majorana masses\cite{zee1}. Some variations on this model\cite
{zee2,zee3} were considered a few years later. The various versions of
this model has been studied over the years,
\cite{wolfenstein,babu,petcov,chang} and 
recently has attracted a great deal of
attention.\cite{smirnov,jarlskog,frampton,joshipura,kim} The purpose of this
paper is to make some further remarks on this class of models. We start with
a review of the model and its variations, so that this paper is largely self
contained.

\bigskip In the standard model, the neutrinos are of course massless. We
will follow the old fashioned practice of economy, and try to add as
little
theoretical structure as possible. We can introduce either a Dirac mass or a
Majorana mass for the neutrino. In the first alternative, one needs to
introduce right handed neutrino fields and the question immediately arises
on why the neutrino Dirac masses are so small compared to the charged lepton
masses in the theory. This question was answered elegantly by the see-saw
mechanism, in which the right handed neutrino fields are given large
Majorana masses. But if we are willing to introduce Majorana masses to the
right handed neutrino fields, perhaps we should consider dispensing with
right handed neutrino fields altogether and simply try to generate 
Majorana masses for the existing left handed neutrino fields.  We shall
try to generate this Majorana masses through quantum mechanical effect.
This has the added advantage of having naturally small neutrino mass
instead of relying on a new heavy scale for this explanation in the seesaw
scheme.

Since in the standard model, the left handed neutrino fields belong to
doublets $\psi_{aL}=\left(
\begin{array}{c}
\nu _{a} \\
l_{a}
\end{array}
\right) _{L}$ (with $a$ a family index) we cannot simply put in Majorana
mass terms. The general philosophy followed in Ref.\cite{zee1,zee2,zee3} is
that we should feel freer to alter the scalar field sector than other
sectors since the scalar field sector is the least established one in the
standard model. Out of the doublets we can form the Lorentz scalar $(\psi
_{aL}^{i}C\psi _{bL}^{j})$ (where $i, j$ denote electroweak $SU(2)$
indices and $C$ the charge conjugation matrix): this can be either a
triplet or a singlet under $SU(2)$.  If we couple a triplet field to this
lepton bilinear,
then when the neutral component of the triplet field acquires a vacuum
expectation value, the neutrinos immediately acquire Majorana masses.  We
considered this model unattractive: not only does it lack predictive
power,
but the rather accurately studied ratio of $W$ and $Z$ boson masses puts 
a stringent bound on any triplet Higgs.  In addition, there is no natural
way to explain the smallness required of this vacuum expectation value.  
We thus chose the alternative of
coupling to an $SU(2)$ singlet (charged) field $h^{+}$ via the term
$f^{ab}(\psi_{aL}^{i}C\psi _{bL}^{j})\varepsilon _{ij}h^{+}.$

An interesting point is that due to Fermi statistics the coupling $f^{ab}$
must be anti-symmetric in $a$ and $b.$ We are forced to couple leptons in
one family to leptons in another one. Thus, the term above contains $%
f^{e\mu }(\nu _{e}C\mu ^{-}-e^{-}C\nu _{\mu })h^{+},$ for instance$.$ As we
will see, this leads to an interesting texture in the resulting neutrino
mass matrix.

The term $f^{ab}(\psi _{aL}^{i}C\psi _{bL}^{j})\varepsilon _{ij}h^{+}$ in
itself does not violate lepton number $L$ since we can always assign $L=-2$
to $h^{+}.$ But we note that we can also couple $h^{+}$ to the Higgs
doublets via $M_{\alpha \beta }\phi _{\alpha }\phi _{\beta }h^{+}$ if there
are more than one Higgs doublet. By Bose statistics, the coupling matrix $%
M_{\alpha \beta }$ is antisymmetric.  If the doublets are required to have
zero lepton number by their respective Yukawa couplings, lepton number is
now violated by two units, just right for generating neutrino Majorana
masses.

\bigskip We do not regard the necessity of more than one Higgs doublet as
unattractive. Indeed, theorists have always been motivated by one reason or
another to introduce additional Higgs doublets. For example, some of the
more
attractive theories of CP violation \cite{weinberg lee} requires more than
one Higgs doublets.

We refer to this class of models as $\{h\phi_1\phi_2\}$. From general
principles we know that neutrino Majorana masses must be generated and that
they must come out as finite, that is calculable in terms of the parameters
of the theory. Indeed, we see that calculable neutrino Majorana masses are
generated by the one loop diagram in Fig.(1).

To see finiteness, one way is to simply count the number of propagators.
Another way, as is well-known, is to note that in this model, $h^{-}$ and
the negatively charged components of $\phi_{1}$ and $\phi _{2}$ mix.  After
one component is eaten by the $W^{-}$ the contribution of the other two
components to the one loop diagram in Fig.(1)
cancel in a scalar version of the Glashow-Iliopoulos-Maiani mechanism.

Whenever more than one Higgs doublets are present, we have to worry about
flavor changing effects from Higgs exchange. In the literature, there is of
course no lack of mechanism to suppress this flavor changing effects. In
the present context, a particularly clean proposal is that of Wolfenstein
\cite{wolfenstein}, who imposed a discrete symmetry so that one of the
two Higgs
doublets, say $\phi _{2}$, does not couple to leptons. By assigning $L=2$ to
$\phi _{2}$ we conserve lepton number. At this point, there are two
possibilities. We can break lepton number softly by a term $m_{12}^{2}\phi
_{1}^{\dagger }\phi _{2}+H.c$. or we can break lepton number spontaneously
by having $\phi _{2}$ acquire a vacuum expectation value and create a
Majoron. Since the doublet Majoron model is already ruled out by LEP data,
we will assume the symmetry is softly broken.

In any case, we will suppose that some unspecified mechanism suppresses
flavor changing effects from Higgs exchange, so that the effective Higgs
couplings in Fig.(1) do not change flavor.  Then we
see that this one-loop diagram gives a neutrino mass matrix with the
texture
\begin{equation}
(m_{\nu })_{ab}=cf_{ab}(m_{a}^{2}-m_{b}^{2})=c([f,m^{2}])_{ab},
\label{1loop}
\end{equation}
where $c$ is $a_{1}(M_{12}/M_{h}^{2})\ln (m_{\phi }^{2}/M_{h}^{2})$ and $%
a_{1}$ is of order of the one-loop factor $1/(16\pi ^{2})$. In
particular, the
diagonal elements vanish. Various authors\cite{jarlskog,frampton} have
used this texture to fit the atmospheric and solar neutrino data (but not
the LSND data). 
They found that the data can be accommodated if 
$f_{\mu\tau} << f_{e\tau} << f_{e\mu}$, and 
$f_{e\tau}m_{\tau}^{2} \sim f_{e\mu}m_{\mu}^{2}$.
In that case, the solar neutrino oscillation would be due to either
large angle MSW or vacuum oscillation.  The neutrino mixing matrix has the
form 
$$
U^{\nu}=\left(
\begin{array}{ccc}
{1 \over \sqrt{2}} & {1 \over \sqrt{2}} & 0 \\
-{1 \over 2} & {1 \over 2} & {1 \over \sqrt{2}} \\
-{1 \over 2} & {1 \over 2} & {1 \over \sqrt{2}}
\end{array}
\right) 
$$ 
and $c f_{e\mu} m^2_\mu = c f_{e\tau} m^2_\tau = \sqrt{\Delta M^2_{atm}/2}
= (1.6 $--$ 5.5)\times 10^{-2} eV$.  The coupling $f_{\mu\tau}$ can be
related to $f_{e\tau}$ through $f_{\mu\tau} = f_{e\tau}(\Delta
M^2_{solar}/\Delta M^2_{atm})$.  According to the recent data
\cite{superk} $\Delta M^2_{solar} = 1.8 \times 10^{-5} eV^2$
with $\sin^2\theta= 0.76$ if one take the large angle MSW solution while 
$\Delta M^2_{solar} = 6.5 \times 10^{-11} eV^2$ with $\sin^2\theta= 0.75$
for
vacuum oscillation solution.  The same experiment also found
$\Delta M^2_{atm} = (0.5 $--$ 6) \times 10^{-3} eV^2$ with $\sin^2\theta >
.82)$ for atmospheric neutrino oscillation.
 
The observed hierarchy of couplings 
( $f_{\mu \tau} << f_{e\tau} << f_{e\mu}$)
may indicate the approximate conservation of the additive quantum number $%
L_{\mu }+L_{\tau }-L_{e},$ as we will discuss below.

It is worth emphasizing that the texture in (\ref{1loop}) is the result of
the model plus our assumption that flavor-changing Higgs couplings are
suppressed. If we relax the latter assumption, then $(m_{\nu})_{ab}$ would
be a general $3$ by $3$ symmetric matrix.

The phenomenological consequences of introducing the field $h$ have been
well studied\cite{zee1,zee2,ng} and so we will not go into it here.  We
will merely note one particularly attractive feature\cite{zee2}.  Probably
one of the best measured processes in leptonic weak interatcion is muon
decay $\mu ^{-}\rightarrow e^{-}+\overline{\nu }_{e}+\nu _{\mu }$\ to which $%
h$ exchange contributes at tree level. Fortunately, $h$ couples to
left-handed fields and so upon Fierz rearrangement we see that only the
overall magnitude, but not the angular distribution, of muon decay is
affected. This sets a relatively weak bound on $f^{2}/m_{h}^{2}.$

Some years after this model was proposed, one of us\cite{zee3} noted that in
line with the same philosophy we can also introduce a doubly charged field $%
k^{++}$ coupling to the right-handed lepton singlets, thus $\tilde{f}%
_{ab}l_{aR}Cl_{bR}k^{++}.$ Fermi statistics now requires that $\tilde{f}$ be
symmetric. Lepton number is broken by the trilinear coupling $\tilde{M}
k^{++}h^{-}h^{-}$. One therefore does not need to introduce additional
doublet.  The two-loop diagram in Fig.(2) generates a
two-loop contribution to the neutrino Majorana matrix of the form
\begin{equation}
(m_{\nu }^{(2)})_{ab}=c_{2}\sum_{cd} f_{ac}\tilde{f}_{cd}f_{db}
\tilde{M}(m_{c}m_{d}/M_{k}^{2})  \label{2loop}
\end{equation}
where $c_{2}$ is a two-loop factor 
$c_{2}=[ln(M_{k}^{2}/M_{h}^{2}+1)]^{2}/(32\pi ^{4})$\cite{babu}. 
We refer to this as the $\{h\phi k\}$ model. 
One of course can also have additional doublets and the associated one
loop contribution to neutrino mass.
We refer to this as the $\{h\phi _{1}\phi _{2}k\}$ model.  In this case
the one-loop and the two-loop contributions are to be added.


Later, Babu\cite{babu} studied these two classes of models. In particular,
in the $\{h \phi k\}$ model we have $Det[m^{(2)}]=0$ since $Det[f]=0$ due to
the fact that $f$ is a $3\times 3$ antisymmetric matrix.

At this point we depart from our general treatment and focus on specific
possibilities as suggested by experiments. The various possible textures of
the Majorana neutrino mass matrix, in particular those dictated by the
conservation of additive combination of electron, muon, and tau numbers,
such as $L_{\mu }+L_{\tau }-L_{e},$ were studied in \cite{zee3}. Typically,
it is awkward in many models of neutrino masses to impose conservation of
these additive combinations, however, as was pointed out in \cite{zee3}, 
it can be quite
naturally implemented in this class of models by simply setting various
couplings to zero. For example, suppose we set $f_{\mu \tau }$ to zero. Then
the $f_{e\mu }$ term demands that the $h$ field to carry $L_{e}=-1,L_{\mu
}=-1,$ and $L_{\tau }=0,$ while the $f_{e\tau }$ term demands that the $h$
field to carry $L_{e}=-1,L_{\mu }=0,$ and $L_{\tau }=-1.$ The clash between
these two terms implies that $L_{\mu }$ and $L_{\tau }$ are violated, but
that $L_{\mu }+L_{\tau }-L_{e}$ and $L_{e}$ are conserved.

One purpose of this paper is to study the two-loop contribution to $m_{\nu }$
in the $\{h\phi _{1}\phi _{2}\}$ model. The relevant diagrams are shown in
Fig.(3).  For instance, the diagram in Fig.(3a)
contributres to 
$(m_{\nu })_{ab}$ a term
\begin{equation}
(m_{\nu }^{(2)})_{ab}=\gamma
\sum_{c,d}f_{ac}f_{cd}^{*}f_{db}(m_{c}^{2}-m_{d}^{2})=\gamma
(f[m^{2},f^{*}]f)_{ab}  \label{2looporig}
\end{equation}
\bigskip where $\gamma =a_{2}(16\pi ^{2})^{-2}(M_{12}/M_{h}^{2})$ with $%
a_{2} $ of order one. We are interested in the diagonal entries in $(m_{\nu
}^{(2)})_{ab}$ since the off-diagonal entries just give a small perturbation
to the one-loop contribution in (\ref{1loop}). We see from the antisymmetry
of $f$ that the diagonal elements necessarily involve the product of all
three of the non-zero $f_{ab}$, ($a\neq b$). Thus, for instance, $(m_{\nu
}^{(2)})_{ee}=\gamma f_{e\tau }f_{\tau \mu }^{*}f_{\mu e}(m_{\tau
}^{2}-m_{\mu }^{2})\sim \gamma f_{e\tau }f_{\tau \mu }^{*}f_{\mu e}m_{\tau
}^{2}.$ Note that similarly $(m_{\nu }^{(2)})_{\mu \mu }\sim \gamma f_{e\tau
}^{*}f_{\tau \mu }f_{\mu e}m_{\tau }^{2}$ which is equal to $(m_{\nu
}^{(2)})_{ee}.$ An interesting texture emerges upon noting that $(m_{\nu
}^{(2)})_{\tau \tau }$ is smaller by a factor $m_{\mu }^{2}/m_{\tau }^{2}.$

Thus, for phenomenological analysis we have a neutrino mass matrix of the
form 
$$
m_{\nu }=\left(
\begin{array}{ccc}
r & a & b \\
a & s & c \\
b & c & t
\end{array}
\right) 
$$
with the texture $a \sim b >> c > r \sim s >> t$. 
We expect that the terms $r \sim s$ would provide small corrections to the
phenomenological analysis of Jarlskog et al \cite{jarlskog}.  

It is probably premature to consider the effects of CP violation; the
enormous difficulty of measuring CP violation in neutrino oscillations has
been discussed recently \cite{ruju}.  But it may still be interesting to
investigate the potential source of CP violation in various models.  For
the Yukawa coupling of $h^+$, assuming that the charged lepton mass matrix
has been diagonalized, it is possible to redefine the phases of $\psi_a$
such that $f_{ab}$ is a real matrix for three generation case.
Therefore, in the $\{h\phi_1\phi_2\}$
model, additional CP violation has to come from the Yukawa coupling of
$\phi_2$.  If $\phi_2$ does not couple to fermions, due to, say, softly
broken lepton number symmetry, there will be no new explicit CP violating
phase in the theory.  Even the phase in the soft breaking term 
$m_{12}^{2}\phi_{1}^{\dagger }\phi _{2} + H.c$ can be absorbed.  
The story is of course
changed if both $\phi_1, \phi_2$ are allowed to to have Yukawa couplings.
On the other hand, in the $\{h\phi k\}$ model, the phases in the Yukawa
coupling
$\tilde{f}$ can not longer be absorbed.  Therefore, there will be CP
violation in the two loop mass matrix in Eq.(2).

While our discussion is not supersymmetric, however as a mechanism it can
be easily imbedded into a supersymmetric theory\cite{otto}.  In particular
the $h^+$ has exactly the gauge quantum number of a right-handed
electron. Therefore if we allow lepton number violating R-parity
breaking terms, the role of the term
$f^{ab}(\psi_{aL}^{i}C\psi_{bL}^{j})\varepsilon _{ij}h^{+}$
in the $\{h\phi_1 \phi_2\}$ model can be played exactly by part of such
R-parity breaking terms in which $h^+$ is the right-handed slepton and
$\phi_i$ are linear combinations of the Higgs doublet and slepton
doublets.  A large portion of our discussions obviously applies to this
class of models.

It was emphasized\cite{chang} some time ago that in some models even if
the Majorana neutrino masses are naturally small due to their quantum
origin, the associated neutrinoless double beta decay\cite{mohapatra} may
be a tree level effect and thus much larger than one would have naively
guessed based on the small masses.  In fact some of the models discussed
here naturally exhibit this possibility.  In the
$\{h\phi_1\phi_2\}$ model, contrary to the claim in Ref\cite{jarlskog},
there is a potential tree level contribution to the neutrinoless double
beta decay as shown in Fig.(4a) if both $\phi_i$ are allowed to couple to
fermions.  The estimate of the diagram is very similar to the ones in
Ref\cite{bm}.  The scalar $h$ induces a four fermion interaction
$\epsilon_1^{\mu e} (4 G_F/\sqrt{2}) \bar{d}_R u_L \nu_{\mu}^T C^{-1}e^L$,
using the notation in Ref.\cite{bm}, with $\epsilon_1^{\mu e}$ estimated
to be $(4 G_F/\sqrt{2})^{-1} f_{e\mu}h_d v_2 M_{12}/(M_h^2 M_\phi^2)$
where $h_d$ is the Yukawa coupling of the down quark to $\phi_1$.  One
also
needs the exchange of the physical charged Higgs boson $H^+$ in order to
change $\nu_\mu$ back into electron in Fig.(4a).  We parameterize this
$H^+$ exchange amplitude by taking its ratio with the $W$ exchange
amplitude as in 
$r_{e\mu} = (\tilde{h}_{du} \tilde{h}_{e\mu}/M_H^2)(4 G_F/\sqrt{2})^{-1}$
where $\tilde{h}_{ij}$ are the coupling constants of $H^+$.
The current experimental constraint is 
$r_{e\mu} \epsilon_1^{\mu e} < 1 \times 10^{-8}$.  
Taking $cf_{e\mu}m_\mu^2 = 4 \times 10^{-2} eV$ from the neutrino data and
$h_d v_2 \sim m_d \sim 6 MeV$ and $M_\phi = 100 GeV$, one obtains the
limit $r_{e\mu} (1.6 \times 10^{-4}) < 1 \times 10^{-8}$.
This gives rise to stringent constraint on the angle $r_{e\mu}$.
Similar charged Higgs exchange can also give rise to $e-\tau$ flavor
mixing.  The corresponding parameter, $r_{e\tau}$, is constrained to be
$r_{e\tau}(1.6 \times 10^{-4})(m_\mu/m_\tau)^2 < 1 \times 10^{-8}$.
Note that if one uses $W$ exchange instead of $H^+$ exchange, 
the tree level diagram cannot mediate neutrinoless double beta decay. 
However, the diagram in Fig.(4b) can give rise to observable
$\mu^- $--$ e^+$ conversion\cite{dohmen} in nuclei as pointed out in
Ref.\cite{bm}.
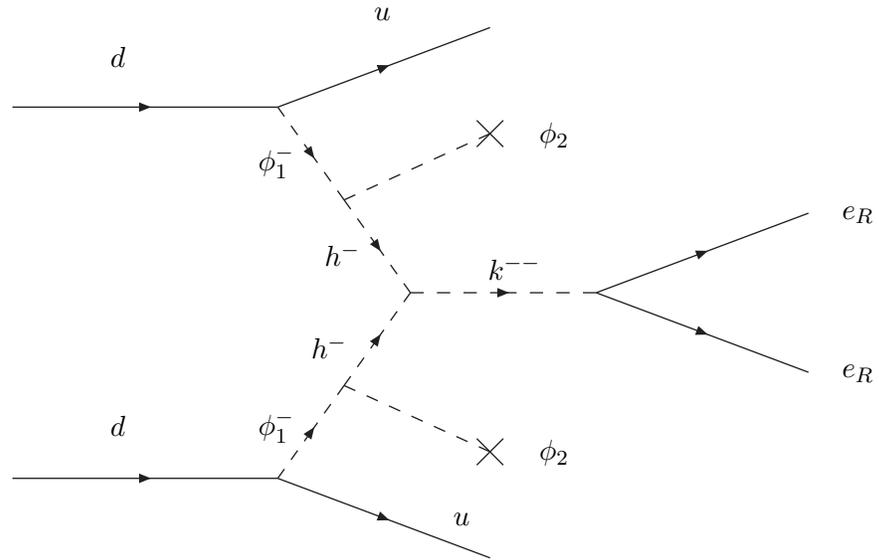
In the $\{h\phi k\}$ model, there is no tree level contribution to
neutrinoless double beta decay.   In the $\{h\phi_1\phi_2 k\}$ model, in
addition to the contribution in Fig.(4a), there are additional tree level
contributions to neutrinoless double beta decay as shown in Fig.(5).
The estimate of this diagram is similar to the one in Ref.\cite{chang}.
The diagram gives rise to the operator $G_2 (\bar{d}_R u_L)^2
e_R^TC^{-1}e_R$ with $G_2 \sim (h_d v_2 M_{12})^2 \tilde{f}_{ee}\tilde{M}
/(M_h^4 M_\phi^4 M_k^2)$.
Assuming that various contributions do not cancel each other, we find that 
the current experimental limit gives rise to the constraint
$|(G_2 m_p/G_F^2)4\pi \lla f|\Omega_\Delta|i \rra| < 0.8 \times 10^{-4}$
where
$\lla f|\Omega_\Delta|i \rra$ is a dimensionless nucleus matrix element as
defined
in Ref.\cite{chang}.  A rough estimate\cite{chang} of $4\pi
\lla f|\Omega_\Delta|i \rra$ is $200$.  For example, taking $M_h \sim
M_\phi \sim M_k \sim
M_{12} \sim \tilde{M} \sim 100 GeV$, this constrains $\tilde{f}_{ee} <
0.16$.

In conclusion, we have summarized some interesting models of
radiative neutrino masses without any additional fermions added to the
standard model.  We emphasized that, in the simplest $\{h\phi_1 \phi_2\}$
model, in addition to the one loop diagram, there are two loop diagrams
that may give rise to interesting texture of neutrino mass for potential 
future experimental test.  We point out that in some of these models, the
neutrinoless double beta decay may be a tree level effect and, as a
result, may give rise to serious contraint on the theory when combined
with the neutrino oscillation data.  In addition, the model may give rise 
to observable $\mu^- $--$ e^+$ conversion in nucleus.

{\it Acknowledgements : }
AZ would like to thank the staff of the National Center for Theoretical
Sciences where this work was done for its warm hospitality. We are grateful
to the National Center for Theoretical Sciences of National Science
Council(NSC) of Republic of China for support which makes this research
possible.  
DC is supported by a grant from the NSC of ROC and 
AZ by a grant from the NSF of USA.


\begin{references}
\bibitem{superk}  Y. Fukuda et. al., Phys. Lett. {\bf 436B} 33(1998); Phys.
Rev. Lett. {\bf 81} 1158(1998); Phys. Rev. Lett. {\bf 81} 1162(1998); S.
Hatakeyama et. al., Phys. Rev. Lett. {\bf 81} 2016(1998);

\bibitem{zee1}  A. Zee, Phys. Lett. {\bf 93B} 389(1980).

\bibitem{zee2}  A. Zee, Phys. Lett. {\bf 161B} 141(1985).

\bibitem{zee3}  A. Zee, Nucl. Phys. {\bf 264B} 99(1986).

\bibitem{wolfenstein}  L. Wolfenstein, Nucl. Phys. {\bf B175} 93(1980).

\bibitem{petcov}  S. T. Petcov, Phys. Lett. {\bf 115B} 401(1982).

\bibitem{babu}  K. S. Babu, Phys. Lett. {\bf B203} 132(1988).

\bibitem{chang}  D. Chang and W.-Y. Keung, Phys. Rev. {\bf D39} 1386(1989).

\bibitem{ng}  G. C. McLaughlin and J.N. Ng hep-ph/9907449.

\bibitem{smirnov}  A. Yu Smirnov and M. Tanimoto, Phys. Rev. {\bf D55}
1665(1997)

\bibitem{jarlskog}  C. Jarlskog, M. Matsuda, S. Skadhauge and M. Tanimoto,
Phys. Lett. {\bf B449}, 240(1999).

\bibitem{frampton}  P. H. Frampton and S. L. Glashow, hep-ph/9906375.

\bibitem{joshipura}  A. S. Joshipura and S. D. Rindani, hep-ph/9907390.

\bibitem{kim}  J. E. Kim and J. S. Lee, hep-ph/9907452.

\bibitem{weinberg lee}  S. Weinberg, Phys. Rev. Lett. {\bf 37}, 657(1999); 
T. D. Lee, Phys. Rep. {\bf 9}, 143(1974).

\bibitem{otto} O. Kong, Mod. Phys. Lett. {\bf A14}, 903(1999); L. Clavelli
and P. H. Frampton, hep-ph/9811326.

\bibitem{ruju}  A. deRujula et al, hep-ph/9811390v2

\bibitem{mohapatra}  For a review, see R. N. Mohapatra, hep-ph/9507234.

\bibitem{bm} K. S. Babu and R. N. Mohapatra, Phys. Rev. Lett. {\bf 75}
2276(1995).

\bibitem{dohmen} C. Dohmen et. al., Phys. Lett. {\bf B317} 631(1993).

\end{references}
\end{document}